\documentclass{report}

\usepackage{hyperref}
\hypersetup{
    colorlinks,%
    citecolor=blue,%
    filecolor=blue,%
    linkcolor=blue,%
    urlcolor=blue,
}  
 
\usepackage{natbib}

\usepackage{listings}
\usepackage{color}

\usepackage{amsmath}
 
\definecolor{codegreen}{rgb}{0,0.6,0}
\definecolor{codegray}{rgb}{0,0,0.5}
\definecolor{codepurple}{rgb}{0.58,0,0.82}
\definecolor{backcolour}{rgb}{0.99,0.99,0.99}
 
\lstdefinestyle{mystyle}{
    backgroundcolor=\color{backcolour},   
    commentstyle=\color{codegreen},
    keywordstyle=\color{magenta},
    numberstyle=\tiny\color{codegray},
    stringstyle=\color{codepurple},
    basicstyle=\footnotesize,
    breakatwhitespace=false,         
    breaklines=true,                 
    captionpos=b,                    
    keepspaces=true,                 
    numbers=left,                    
    numbersep=5pt,                  
    showspaces=false,                
    showstringspaces=false,
    showtabs=false,                  
    tabsize=2
}
\usepackage[caption = false]{subfig}

\usepackage{float}

\usepackage{algpseudocode}

\usepackage{graphicx}

\usepackage{tabularx}

\makeatletter
\def\BState{\State\hskip-\ALG@thistlm}
\makeatother

\graphicspath{{./Figures/}}

\usepackage{url}

\usepackage{fancyhdr}
\usepackage{fancyhdr}
\usepackage{lastpage}

\usepackage{amsmath}
\usepackage{amssymb}
\usepackage{graphicx}
\usepackage{epstopdf}
\usepackage{inputenc}
\usepackage{geometry}
\title{%
  \Huge OSmOSE report $No$ 2 \\ 
  \LARGE Development details and \\ computational benchmarking of DEPAM}

\date{}

\begin{document}

\maketitle
\begin{abstract} 
In the big data era of observational oceanography, passive acoustics datasets are becoming too high volume to be processed on local computers due to their processor and memory limitations. As a result there is a current need for our community to turn to cloud-based distributed computing. We present a scalable computing system for FFT (Fast Fourier Transform)-based features (e.g., Power Spectral Density) based on the Apache distributed frameworks Hadoop and Spark. These features are at the core of many different types of acoustic analysis where the need of processing data at scale with speed is evident, e.g. serving as long-term averaged learning representations of soundscapes to identify periods of acoustic interest. In addition to provide a complete description of our system implementation, we also performed a computational benchmark comparing our system to three other Scala-only, Matlab and Python based systems in standalone executions, and evaluated its scalability using the speed up metric. Our current results are very promising in terms of computational performance, as we show that our proposed Hadoop/Spark system performs reasonably well on a single node setup comparatively to state-of-the-art processing tools used by the PAM community, and that it could also fully leverage more intensive cluster resources with a almost-linear scalability behaviour above a certain dataset volume. 
\end{abstract}

\newpage

\paragraph{Authorship}

This document was drafted by Paul Nguyen (Sorbonne Universit\'es) and Dorian Cazau (Institute Mines Telecom Atlantique, corresponding author, m: cazaudorian@outlook.fr).

\paragraph{Document Review}

Though the views in this document are those of the authors, it was reviewed by a panel of experts, listed below, before publication on arxiv. This enabled a degree of consensus to be developed with regard to the contents, although complete unanimity of opinion is inevitably difficult to achieve. Note that the members of the review panel and their employing organisations have no liability for the contents of this document. The Review Panel consisted of the following experts (listed in alphabetical order):

\begin{itemize}
\item Joseph Allemandou$^3$ 
\item Dorian Cazau$^1$ 
\item Alexandre Degurse$^4$ 
\item Paul Nguyen Hong Duc$^2$
\end{itemize}

belonging to the following institutes (at the time of their contribution)
\begin{enumerate}
\item Lab-STICC, Institute Mines Telecom Atlantique
\item Institut Jean le Rond d'Alembert, Sorbonne Universit\'es
\item JoalTech
\item ENSTA Bretagne
\end{enumerate}

\paragraph{Last date of modifications}

\today

\paragraph{Recommended citation} "Development details and computational benchmarking of DEPAM", OSmOSE report $No$ 2, arXiv:1903.06695

\paragraph{Future revisions} Revisions to this guide will be considered at any time. Any suggestions for additional material or modification to existing material are welcome, and should be communicated to Dorian Cazau (cazaudorian@outlook.fr).

\paragraph{Document and code availability}

This document has been made open source under a Creative Commons Attribution-Noncommercial-ShareAlike license (CC BY-NC-SA 4.0) on arxiv (arXiv:1903.06695). All associated codes have also been released in open source and access under a GNU General Public License and are available on github (\url{https://github.com/Project-ODE/FeatureEngine}).

\paragraph{Acknowledgements}

All codes have been produced by the developer team of the non-profit association Ocean Data Explorer (w: \url{https://oceandataexplorer.org/}, m: oceandataexplorer@gmail.com). We thank the P\^ole de Calcul et de Donn\'ees Marines (PCDM) for providing DATARMOR storage, data access, computational resources, visualization, web-services, consultation,  support services (http://www.ifremer.fr/pcdm). The authors also would like to acknowledge the assistance of the review panel, and the many people who volunteered valuable comments on the draft at the consultation phase.

\paragraph{Material used in other publications}

The content of this document has been partially used in the following publications: 

\begin{itemize}
\item Nguyen, P., Degurse, A., Allemandou, J., Adam, O., Gerard, O., White, P. R., Fablet, R. and Cazau, D. (2019) "A scalable Hadoop/Spark framework for general-purpose analysis of high volume passive acoustic data", IEEE OCEANS 2019, Marseille, June 17-20. 
\end{itemize}

\tableofcontents

\newpage

\chapter{Introduction}

\section{Context}

Technological progress in observational oceanography gave rise to a two-tiered system in which major strategic investments have been put primarily in data acquisition rather than in data management and processing plans. As a result, there is currently a huge gap between in-situ small-scale data acquisition and a more  integrated  global knowledge that could be directly used in operational oceanography research and by decision-making managers. A good example of scientific community facing these difficulties is the underwater Passive Acoustic Monitoring (PAM) one, which investigates biological (e.g, whale census) and human (e.g., ship noise monitoring) activities, as well as physical processes (e.g, wind speed and rainfall estimation), in the ocean. Specifically, due to the development of cabled observatories that now provide virtually unlimited power for high bandwidth, continuous data acquisition, and the increase of storage capacity and life battery of temporary recorders, the volume of datasets to process has become larger and larger. For instance, the PerenniAL Acoustic Observatory in the Antarctic Ocean (PALAOA) observatory has been recording quasi-continuously the underwater soundscape of the Southern Ocean since 2005 \cite{Boebel2006b}, generating about 140 GB per day \cite{Kindermann2008}, and the Ocean Network Canada has collected more than 300 TB of PAM data in their database \cite{Biffard2018}. In France, governmental agencies like Service Hydrographique et Oc\'eanographique de la Marine (SHOM) and Agence Fran\c caise de la Biodiversit\'e (AFB) are also experiencing similar challenges of processing large volume of data in the Directive Cadre Strat\'egie pour le Milieu Marin (DCSMM) context, where anthropogenic ambient noise analysis and marine mammal census have to be performed on a long-term continuous effort. 

Several projects have started to address the question of processing high volume PAM data more efficiently by adopting distributed computing systems. A distributed computing system can be simply defined as a ``system whose components are located on different networked computers (or nodes), which communicate and coordinate their actions by passing messages to one another" \footnote{From https://en.wikipedia.org/wiki/Distributed\_computing.}. Each computer has its own multiprocessor structure and memory. This makes it good for redundant storage and availability, durability. In contrast, local systems based on a single node, as usually used in the PAM community, all processors may have access to a shared memory to exchange information between processors, like when performing multiprocessor parallel computing. Among well-known distributed environments within the big data space, Apache Spark has become a prominent player. Initially developed in 2012 at the AMPLab at UC Berkeley, Spark is an ``open-source distributed general-purpose cluster-computing framework, also providing an interface for programming entire clusters with implicit data parallelism and fault tolerance" \footnote{From https://en.wikipedia.org/wiki/Apache\_Spark.}. 

Big data analytics in the PAM community is only in its early stages. Spark Streaming has already been developed for routine processing (filtering and spectrogram analysis) for large terrestrial datasets \cite{Thudumu2016}. Using Hadoop Distributed File System (HDFS) as a distributed storage system, their system resulted in better runtime performance in comparison to standalone execution, with approximately 78\% reduction in the execution time. Concurrent computational approaches have used the Matlab Parallel Computing Toolbox and Matlab Distributed Computing Server to run detection and classification algorithms for whale species recognition \cite{Dugan2016c}. Their most improved process (classifier-based detection) was 6.57× faster for an 8-node server over a serial process. Other more bespoke master-slave models with data distribution have also been developed, where a master node first splits large audio files into smaller chunks, creates a list of work tasks that are distributed over the nodes and eventually aggregates each node output. Such a system has been developed for acoustic event detection and bioacoustic spectral indices \cite{Truskinger2014}, improving average execution time by 24x for a 5 instance, 32 thread distributed cluster over a single threaded process. Using a similar approach optimized for scalability, a pipeline of different preprocessing operations to reduce concurrent noise sources in audio recordings, is 21.76 times faster with 32 cores over 8 virtual machines, compared to a serial process \cite{Brown2018}.

\section{Contributions}

In this paper, we wish to share these efforts by proposing a scalable computation chain for FFT (Fast Fourier Transform)-based features based on the Hadoop and Spark frameworks. These features (e.g., full frequency band Sound Pressure Levels, SPL) are at the core of many different types of acoustic analysis where the need of processing data at scale with speed is evident. For example, these features often serve as long-term averaged learning representations (among them, the well-known Long-Term Spectral Average, LTSA) of soundscapes to identify periods of acoustic interest \cite{Erbe2015,Merchant2015}, which is done either manually or with image-based pattern recognition methods \cite{Frasier2018}. Such applications, namely fast automatic content report and interactive annotation of large datasets, need fast and scalable computations of the features to be performed efficiently. Furthermore, LTSA generation relies on several processing parameters (e.g., analysis window size) that can highly modify event-specific averaged patterns and reduce the interpretability of LTSA \cite{Hawkins2014}. To better assess this variability, systematic comparative testing of different parameter sets need to be carried out, which also requires intensive computing. 

In addition to develop a new scalable Hadoop/Spark system for high performance computing of FFT-features, we also provide a computational benchmark comparing our system against three other computing systems based on different programming languages (Scala, Matlab and Python) in standalone executions, using execution time as evaluation metric. We also evaluate the system scalability in its distributed configuration.

\chapter{Methods}

\section{Proposed Hadoop/Spark-based system}\label{SparkSys}

Our proposed distributed computing system, based on the Apache Hadoop and Spark frameworks, is shown in Figure \ref{archiHadoopSpark}.

\subsection{System overview}

\begin{figure}[htbp]
\centerline{\includegraphics[width=\columnwidth]{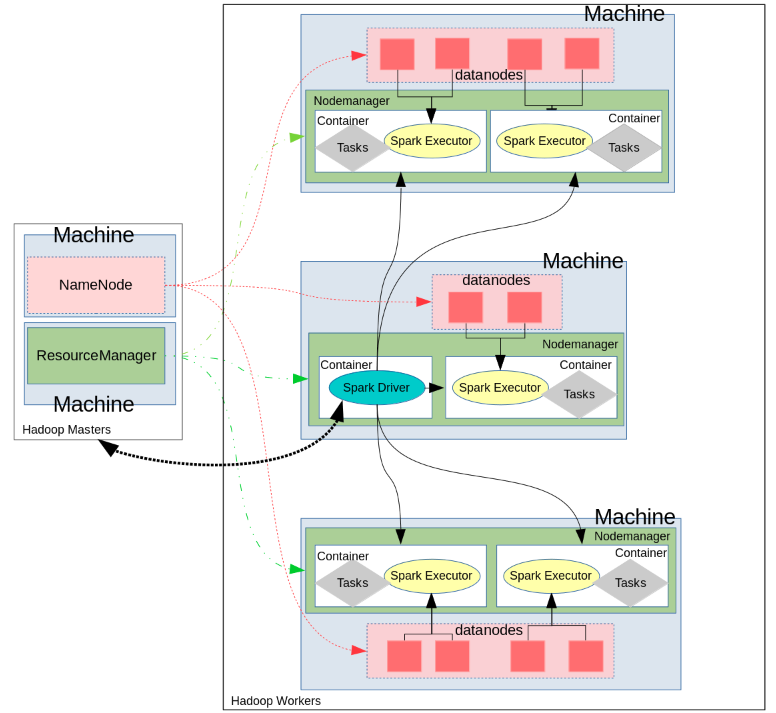}}
\caption{Implementation diagram of the proposed Hadoop/Spark system.}
\label{archiHadoopSpark}
\end{figure}

Hadoop is responsible for distributed data storage and resource management (including job scheduling/monitoring) accross multiple nodes of a cluster, relying respectively on Hadoop Distributed File System (HDFS) and Yet Another Resource Negotiator (YARN). The main function of HDFS is to divide a data file (e.g. 45-mins long / 169 MB for audio files) into smaller blocks of specified size (default is 64 MB). Each block is processed by one map process and map processes run in parallel. In HDFS, the master is the NameNode which manages the filesystem namespace and logs all modifications and the state of the filesystem. It communicates all the information about the content of a filesystem to the DataNodes, corresponding to different machines where the HDFS blocks are locally located. Regarding YARN, the Application Master (here the Spark Driver as we will see after) negotiates resources with the ResourceManager, which is responsible for granting containers corresponding to the resources allocated (container Hadoop = N cores + M GB RAM). Containers are then supervised locally by the NodeManagers. Globally, as represented by the dashed arrows in figure \ref{archiHadoopSpark}, both Hadoop components HDFS and YARN communicate with other machines through a master-slave model as follows: NameNode $\leftrightarrow$ DataNodes for HDFS and ResourceManager $\leftrightarrow$ NodeManagers for YARN. 

Spark focuses on processing data in parallel across a cluster. When used in conjunction with Hadoop, the Spark Driver organizes the completion of the jobs across the cluster of executors by interacting with the ResourceManager and the NodeManagers. Jobs are performed across the worker nodes (CPUs (cores) and allocated memory) using Stages and Tasks. Two main components of the Spark architecture are:

\begin{itemize}
\item Spark Driver: this is the Application Master in our workflow. It tracks all the operations by executors. Moreover, this entity parses the code, and serializes the byte level code across the executors. Any computation is actually done at the local level by each of them. Furthermore, the Driver aims to plan all the computation in the cluster with Directed Acyclic Graph (DAG). Once a DAG is created, it represents a job which is divided in stages. Then, each stage is carried out as tasks. Finally, the Driver handles fault tolerance of all performed operations;

\item Spark Executors: represent processes running in the containers in a cluster. One or more executors could be in each worker node and multiple tasks can be run in a single executor. 
\end{itemize}

\subsection{Implementation details}\label{}

The three parameters \textit{num-executors} (number of executors requested), \textit{executor-cores} (number of tasks the executor can run simultaneously), and \textit{executor-memory} (controlling the executor heap size) play a key role in performance of the Spark system as they control the amount of CPU and memory the application gets. The parameters of the nominal configuration, called SparkSys, have been set as follows: \textit{executor-cores=3}, \textit{num-executors=8} and \textit{executor-memory=11.5GB}. A more optimized version, which corresponds roughly to the best parameter setup in the balance between system performance and resource allocation, and called SparkSysOpti, have also been tested, with parameters: \textit{executor-cores=3}, \textit{num-executors=17} and \textit{executor-memory=5.5GB}. In particular, it was observed that HDFS client has trouble with a great number of concurrent threads, and that it achieves full write throughput with 5 or less tasks per executor.


When using N node clusters, one node is used as a master and remaining N-1 as slaves. Indeed, when a Spark application is run using a cluster manager like YARN, several daemons will run in the background like NameNode, Secondary NameNode, DataNode, JobTracker and TaskTracker. Thus, while specifying \textit{num-executors}, we need to make sure that we leave aside enough cores (typically 1 core per node) for these daemons to run smoothly. Furthermore, the programming language Scala (version 2.11.8) has been used to implement the tested workflow (described in Sec. \ref{WorkflowFeatureDescription}), and we also used the multithreaded FFT library JTransforms \footnote{https://github.com/wendykierp/JTransforms}, written in Java. Note that Scala is a programming language that has flexible syntax as compared to other languages like Python or Java, and Apache Spark itself is written in Scala as it is more scalable on Java Virtual Machine.

\section{Experimental setup}

\subsection{Infrastructure}

All our numerical experimentations have been performed on the DATARMOR infrastructure (\url{http://www.ifremer.fr/pcdm}), belonging to IFREMER. Each node is composed of an Intel Xeon 2X CPU E5-2680 v4 (28c / 56t), 128 GB DDR3, i.e. up to 56 cores (28 $\times$ 2 hyperthreaded CPUs) and 128 GB RAM per node. The multi-node Hadoop-Spark cluster of the SparkSys system was also deployed on DATARMOR. Up to 16 nodes were used to test different distributed configurations o this system. Each node of the cluster runs recent versions of Hadoop and Spark, i.e. 2.8 and 2.4.0, respectively, within the SUSE environment. 

Note that the infrastructure architecture of Datarmor may be under-optimal for computational performance based on Hadoop/Spark frameworks, especially for high volume data, as each node does not have its own hard drive, making the data pass through different pipes with limited I/O throughput instead of being read and written locally. Furthermore, there is no dedicated access to node resources, which are instead shared among users (especially in terms of I/O throughput).

\subsection{Workflow \& dataset}\label{WorkflowFeatureDescription}

The workflow used for the FFT-based feature computations is based on classical PAM analysis blocks (see e.g. \cite{Merchant2015} for background information), including three main steps: short-term FFT analysis (e.g., 32 ms), feature computations and feature integration over longer time segments (e.g., 1 min). Three FFT-based features have been computed: pwelch spectra, Third Octave Levels (TOL) and Sound Pressure Levels (SPL). It is noteworthy that we performed two independent segmentations at different time resolutions to cope with the minimal time window expected in TOL features, set to 1s according to ISO and ANSI standards \cite{ISO1975,ANSI2009}. Also, all the results were sorted by time order and saved in JSON files. A complete description of this workflow, including both theory and implementation details, is available \cite{OSmOSEreport1}. The two parameter sets used for our experiments are listed in table \ref{ParameterSets}.

\begin{table}[htbp]
\centering
\resizebox{0.8\textwidth}{!}{%
\begin{tabular}{|c|c|c|c|}
\hline
Parameter                & Set 1 & Set 2 \\ \hline
nfft            & 256             & 1024           \\ \hline
windowOverlap   & 128             & 0               \\ \hline
windowSize      & 256             & 1024         \\ \hline
recordSizeInSec & 1              & 30           \\ \hline \hline

Processing load &  2700 / 0.06 / 691213     &  90 / 1.92 / 86401           \\ \hline

\end{tabular}
}
\caption{Parameter sets of the FFT-related variables in the workflow: nfft (number of points in FFT), windowOverlap (number of overlapping samples in consecutive windows), windowSize (number of samples of short-term analysis windows), recordSizeInSec (number of samples of longer time segments over which periodograms are averaged). At the line Processing load we report the : number of integration segments per audio file / volume (in MB) of each segment / number of analysis windows per segment.}
\label{ParameterSets}
\end{table}


The dataset used to evaluate computational performance of computing systems is a real underwater PAM dataset recorded at 32,768 Hz near the archipelago of Saint-Pierre-et-Miquelon over the last three months of the year 2010. It consists of 1807 45-min long wav files for a total volume of 320 GB, each file being 169 MB.

\subsection{Tested systems}

Table \ref{ComputingSystems} describes the three different computing systems tested. SparkSys has already been described in Sec. \ref{SparkSys}. It was benchmarked for computational performance against three other systems based on the Scala (v 2.11.8), Matlab (v 2016b) and Python (v 3.5) programming languages, respectively called ScalaSys, MatlabSys and PythonSys.

For all systems, we tried at best to comply with some ``best practices in programming'', drawing from template-like codes that are widely used in the PAM community (e.g., the \textit{PAMGuide} toolbox by \cite{Merchant2015}) for the Matlab implementation, and in the data scientist communities (e.g., the Scipy toolbox, https://www.scipy.org/) for the Scala and Python implementations. Double-precision floating-point format has been used in all three implementations. Multiple unitary tests have been performed on the core features of the workflow, and the outputs were cross-validated with a root mean square error below 10$^{-16}$ \footnote{Cross-validation tests can be reproduced following our codes here https://github.com/Project-ODE/FeatureEngine-benchmark/blob/master/run-tests.sh}.

In their nominal configurations, ScalaSys is the exact same Scala code as in SparkSys but without the Spark and Hadoop connections, using instead parallel collections included in the Scala standard library for multi-threaded processing, set with a default value of 24 threads. MatlabSys runs parallel Matlab code on a single node using the Parallel Computing Toolbox with 24 workers (equivalent to threads). Due to logistic reasons, we have not tested the use of Matlab Distributed Computing Server in a multiple node setup. Also, on a single node, PythonSys uses Scipy as backend to compute acoustic features, and the Multiprocessing library to run computations in parallel. A distributed version of PythonSys based on the Dask library is being developed and will be reported in future publications.

\begin{table}[htbp]
\centering
\resizebox{0.8\textwidth}{!}{%
\begin{tabular}{|c|c|c|c|}
\hline
System                & Language (version) & Parallel framework &  Distributed framework \\ \hline
SparkSys / SparkSysOpti           & Scala (2.11.8)            &  Pseudo-distributed Spark  &  Hadoop / Spark      \\ \hline

ScalaSys            & Scala (2.11.8)            & Standard parallel collections  &  -  \\ \hline

MatlabSys      & Matlab (2016b)           & Parallel Computing Toolbox  &   -    \\ \hline
PythonSys   &   Python (3.5)            & Scipy/Multiprocessing   &    - \\ \hline
\end{tabular}
}
\caption{Nominal configurations of the tested computing systems. The parallel frameworks are used in a single node setup, while the distributed framework is used for a multiple node setup (only tested for SparkSys).}
\label{ComputingSystems}
\end{table}

\subsection{Evaluation}

To evaluate computational performance of our different systems, execution time was assessed. We paid attention that software launch was not included in the execution time computation. Two different types of experimentations have been run, corresponding either to a parallel (single node) or to a distributed (multiple node) setup, as described in the following. We have also performed sensitivity studies for each system individually, which allows to assess performance variability around the nominal configurations given by table \ref{ComputingSystems}. To determine fluctuation in execution times, each tested system configuration has been executed 3 times, then the execution times were averaged over these 3 executions, and the standard deviations were computed. 

\subsubsection{Single node experimentation}

As Spark also supports a pseudo-distributed local mode, we first benchmarked the three computing systems (see table \ref{ComputingSystems}) executed in a single node mode, over a linear increase of workloads, from 0.169 GB (1 wav file) to 16.9 GB (100 wav files).

\subsubsection{Multi-node experimentation}

In a second experimentation, we evaluated the scalability of our Hadoop/Spark system using the speed up metric (also referred to as improvement rate in \cite{Brown2018}), which corresponds to the reduction of execution time due to running a fixed workload using an increased number of hardware processors. A linear increase of workloads, from 16.9 GB (50 wav files) to 270.4 GB (1600 wav files), has been used. Note that, comparatively, workloads used for scalability analysis in the literature appear to be very small, e.g. in \cite{Thudumu2016} and \cite{Brown2018} data volumes of less than 5 GB are used. The other systems have not been tested in their distributed configurations.

\chapter{Results \& Discussion}

\section{Single node experimentation}

\subsection{Benchmarking of systems}

\begin{figure}[htbp]
\centerline{
\includegraphics[width=0.8\columnwidth]{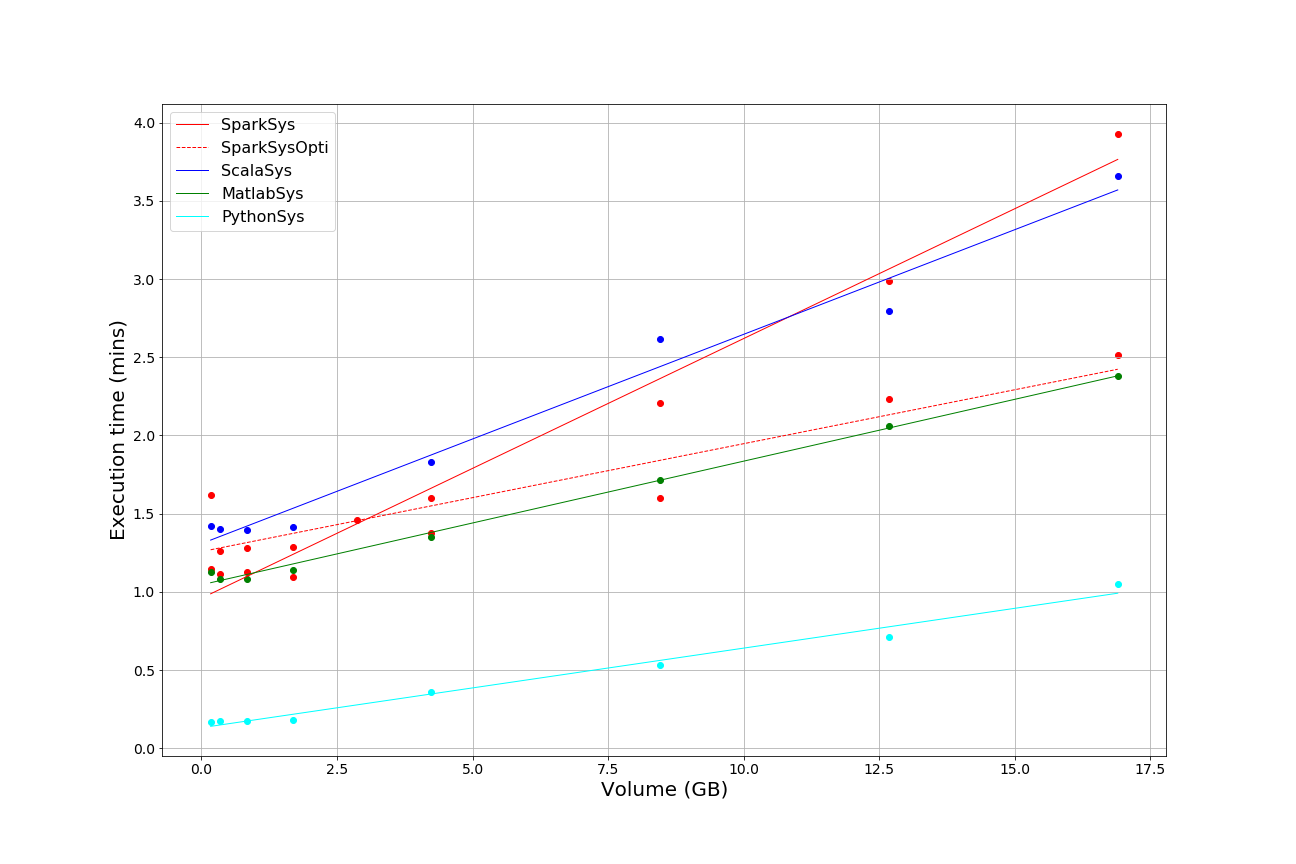}
}
\caption{Execution time (mins) against workload (GB) for parameter set 1.}
\label{singleNodeBestVersions_Set1}
\end{figure}

\begin{figure}[htbp]
\centerline{
\includegraphics[width=0.8\columnwidth]{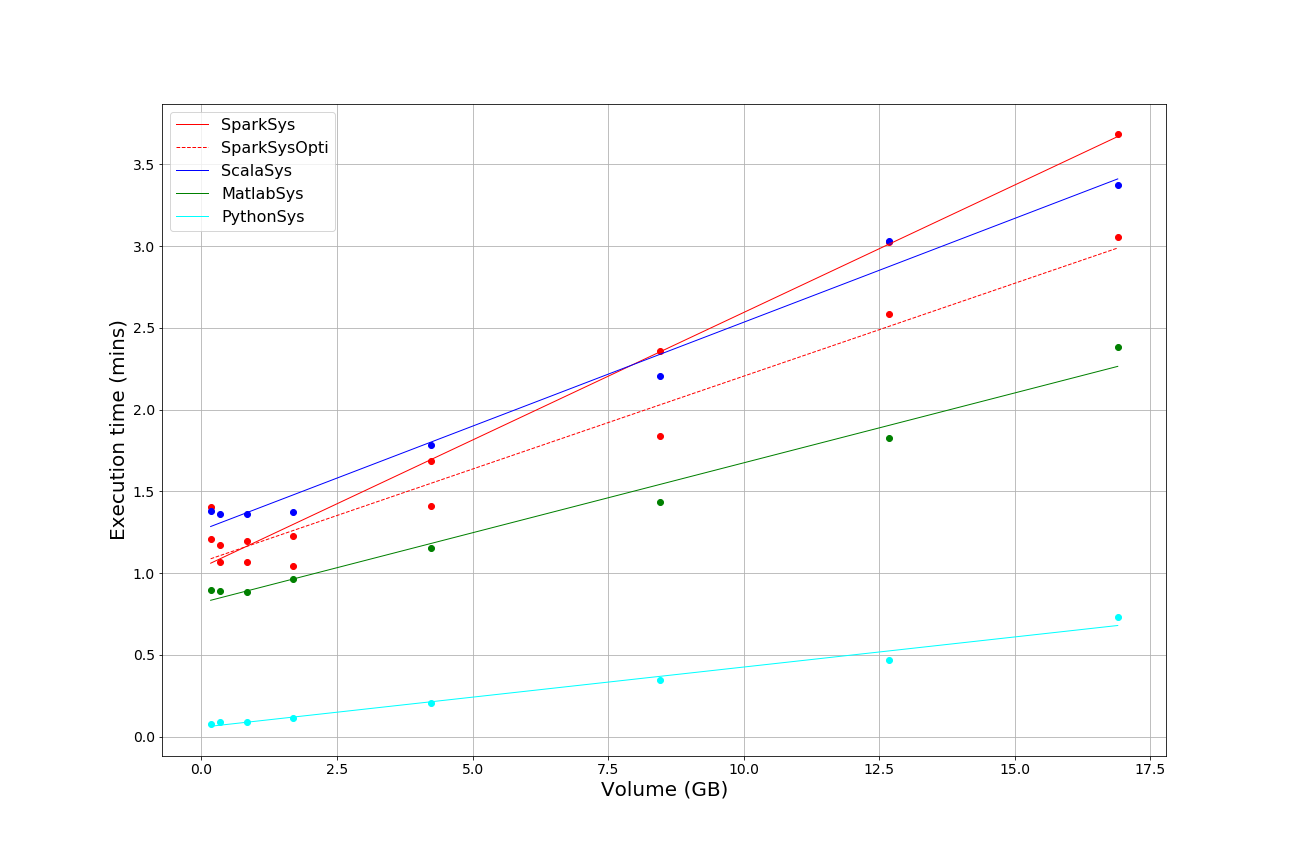}
}
\caption{Execution time (mins) against workload (GB) for parameter set 2.}
\label{singleNodeBestVersions_Set2}
\end{figure}

Figures \ref{singleNodeBestVersions_Set1} and \ref{singleNodeBestVersions_Set2} compare computational performance of the different computing systems in a single node setup, representing execution time (in mins) against workload (GB) for parameter sets 1 (on the left) and 2 (on the right). SparkSysOpti and MatlabSys perform quite similarly, especially for parameter set 1, both largely outperforming ScalaSys and SparkSys, and outperformed by PythonSys. For example, for parameter set 1 and a workload of 16.9 GB (i.e. 100 wav files of our dataset), it takes 2.5 minutes of computation time for SparkSysOpti and MatlabSys, which is more than twice as faster than ScalaSys and SparkSys but more than twice as slower than PythonSys. The difference between SparkSysOpti and MatlabSys is a bit more pronounced on the second set of parameters, although the performance slopes remain quite similar, while PythonSys still outperforms them. Standard deviation values of execution times are relatively minor comparatively to computational gain: 4 s ($\pm$ 3.2) for SparkSys, 2.5 s ($\pm$ 2.2) for ScalaSys and 8 s ($\pm$ 4.8) for MatlabSys.

In definitive, SparkSys performs reasonably well in its standalone mode, with performance similar to the Scala-only version. The slight optimization of its core parameters done with SparkSysOpt allows it to scale up easily and reach MatlabSys performance. This result is particularly interesting as it reveals, although the expected advantage of Apache Spark technology is to scale out processing over several nodes, that our system is also valuable on a single-node architecture, which is the most common computer architecture within the PAM community.

\begin{figure}[htbp]
\centerline{
\includegraphics[width=0.48\columnwidth]{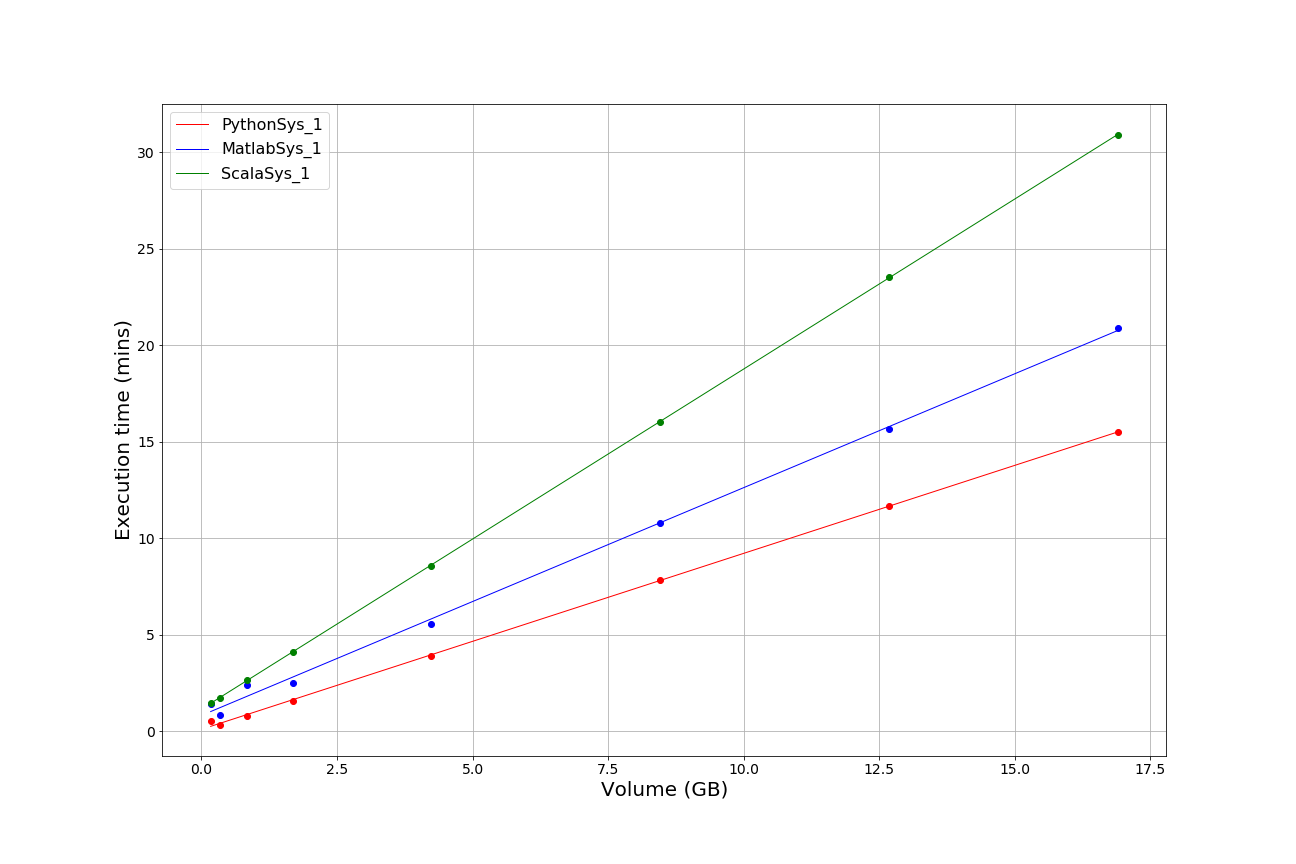}
\includegraphics[width=0.48\columnwidth]{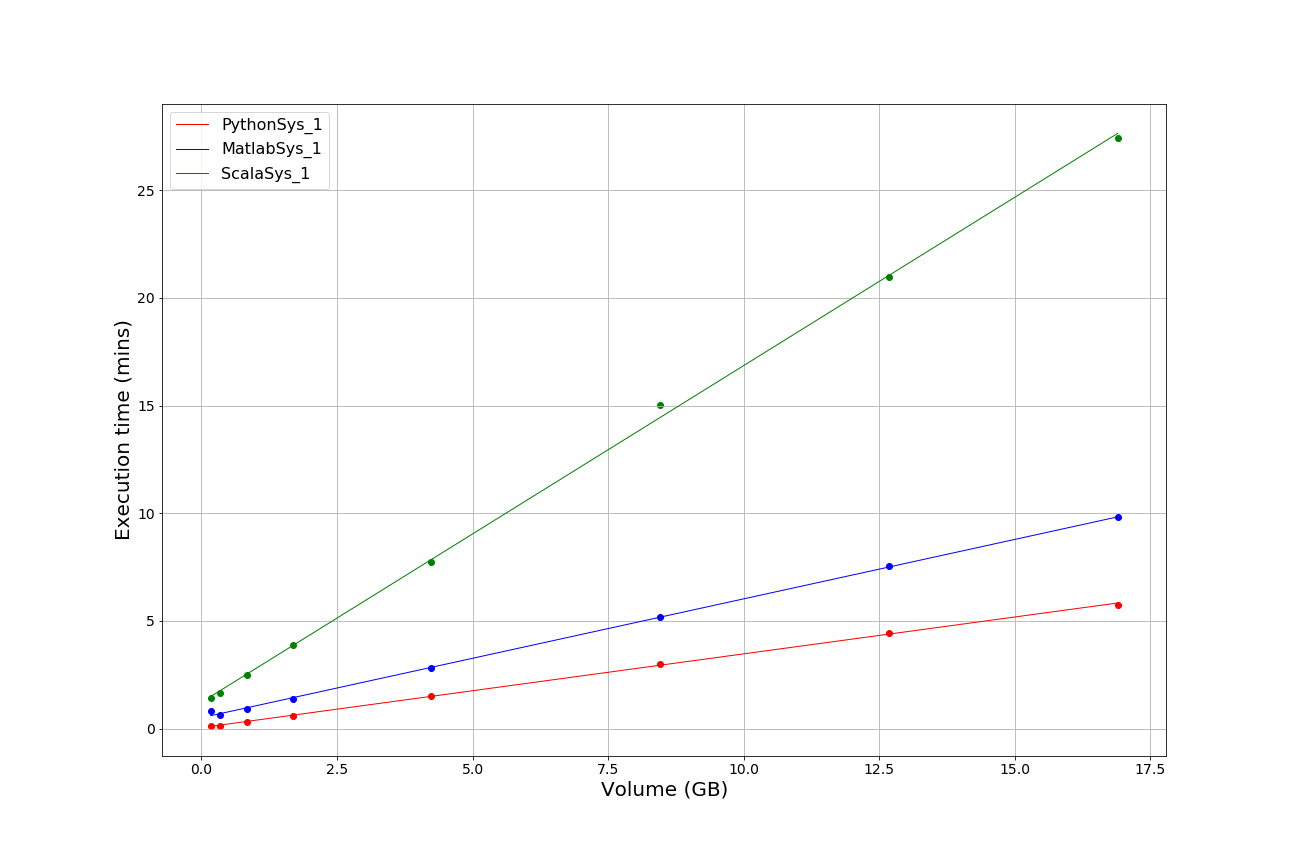}
}
\caption{Execution time (mins) against workload (GB) for parameter sets 1 (on the left) and 2 (on the right).}
\label{singleNode1thread_Set1}
\end{figure}

As a reference, we also provide in figure \ref{singleNode1thread_Set1} the execution times of each version in a single node / single CPU mode, which corresponds to the minimal computer resource we can set. 

\subsection{Sensitivity studies of systems}

\begin{figure*}[t!]
$\begin{array}{rl}
    \includegraphics[width=0.48\textwidth]{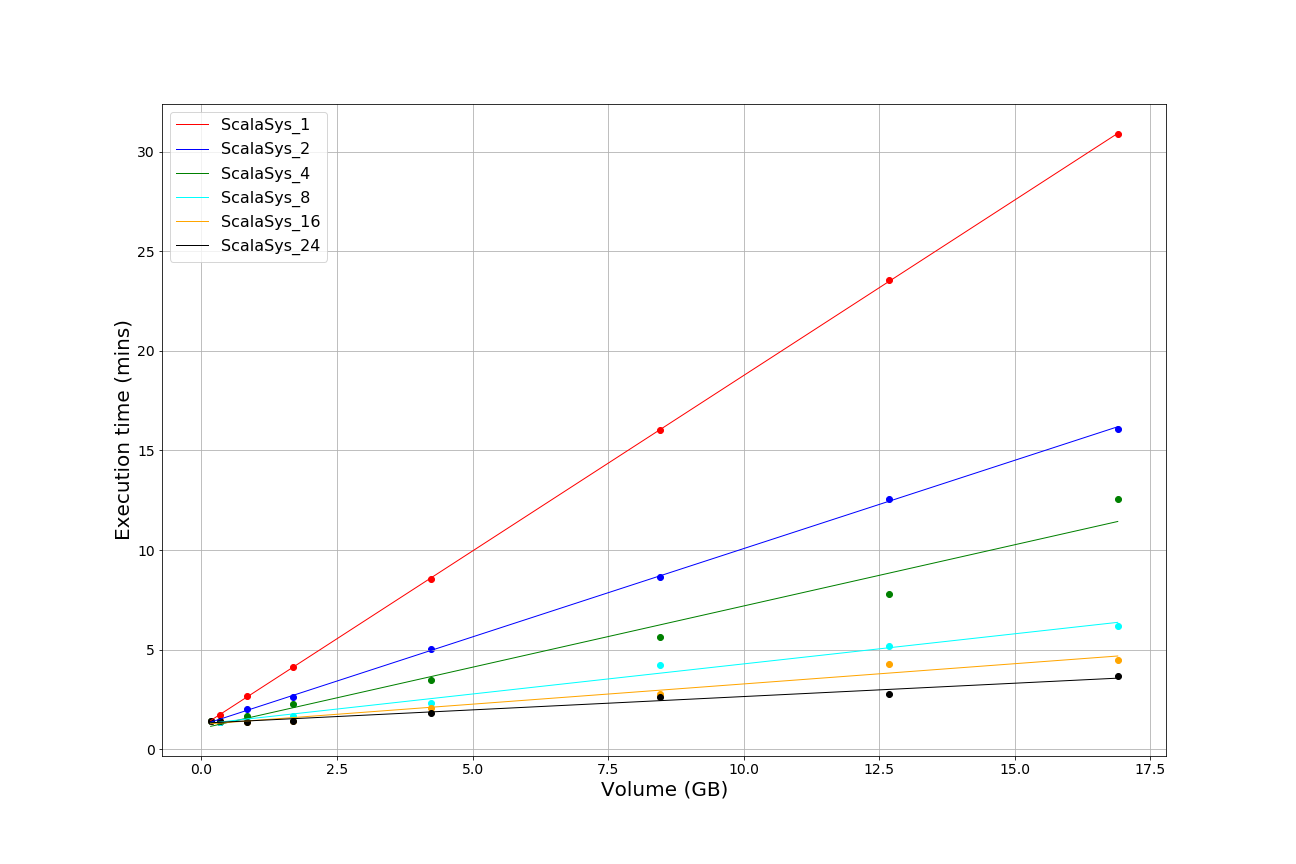} &
    \includegraphics[width=0.48\textwidth]{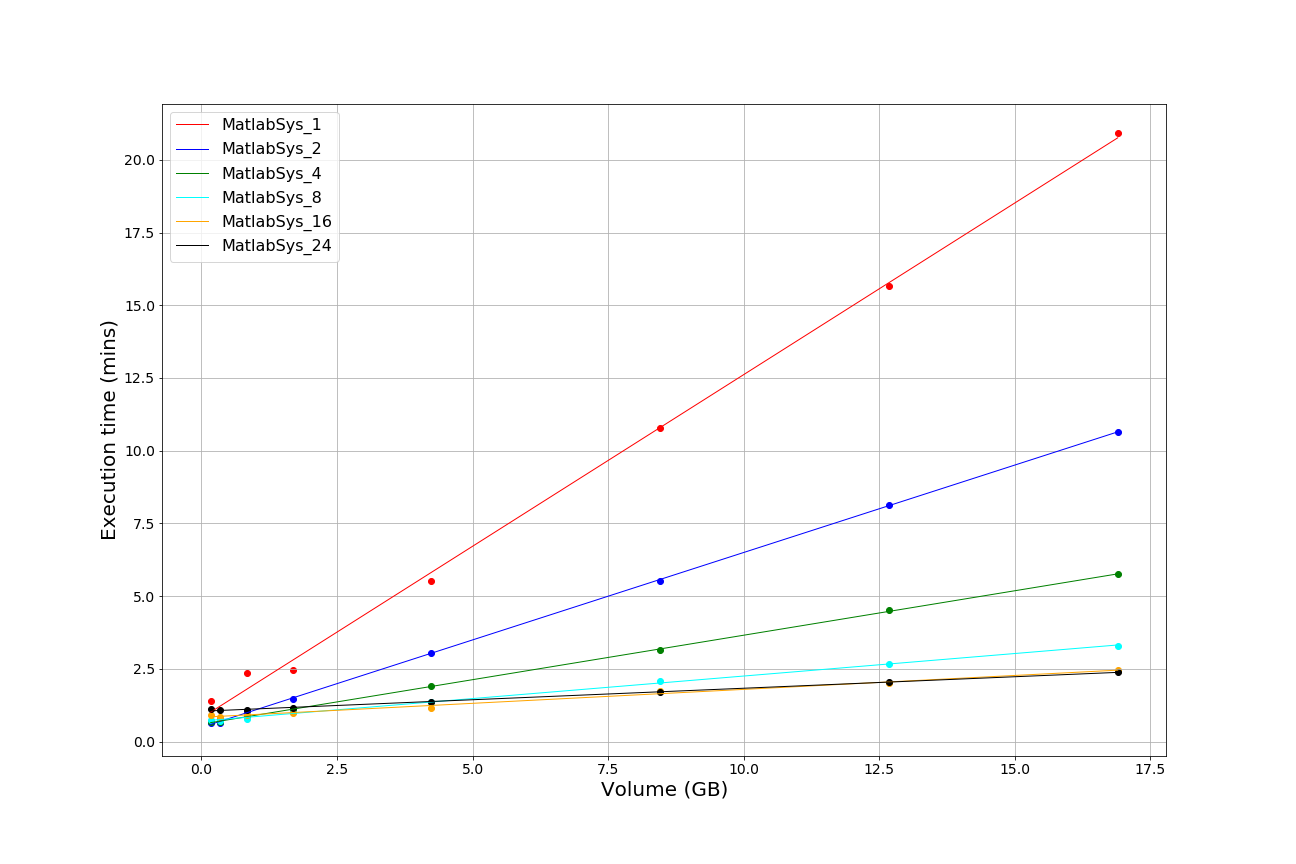}\\
    \multicolumn{2}{c}{\includegraphics[width=0.48\textwidth]{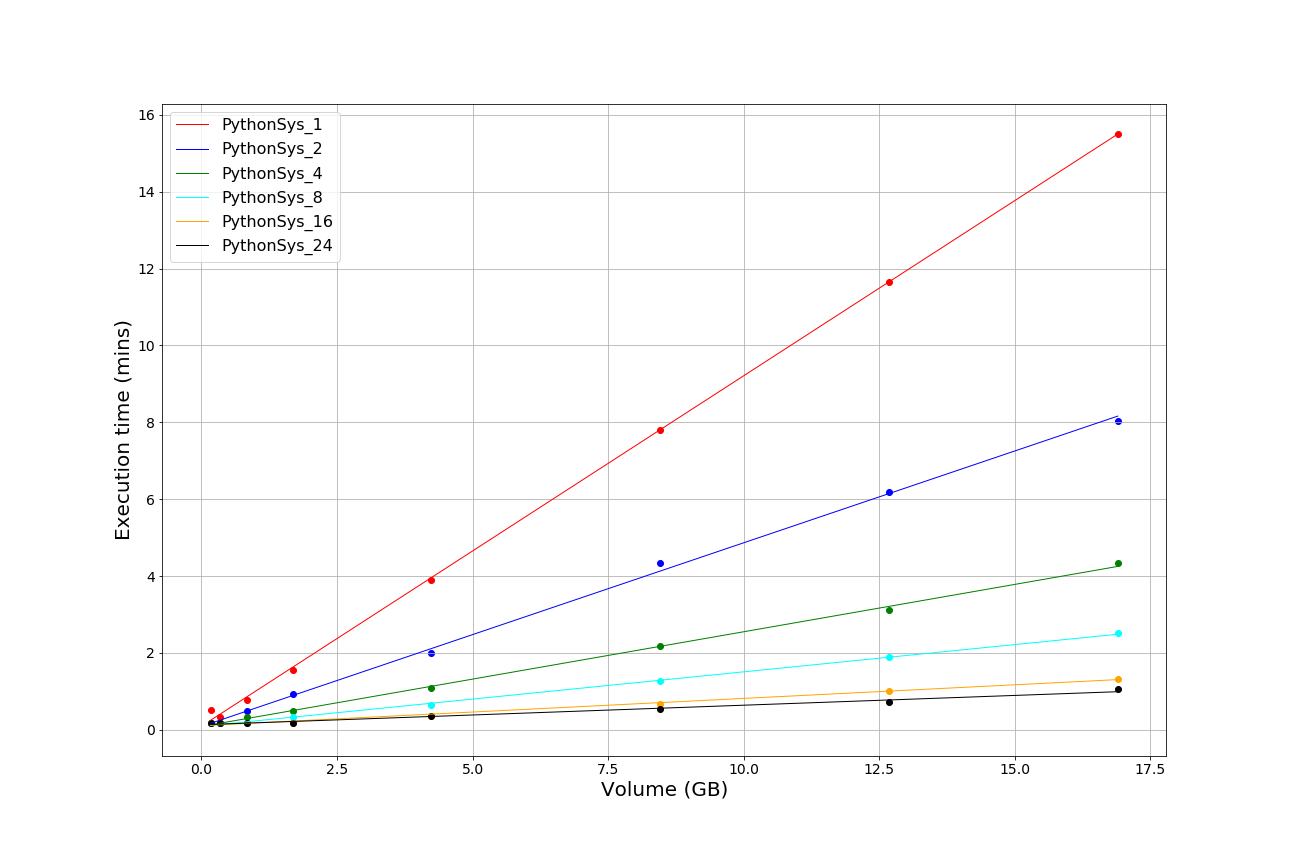}}
\end{array}$
\caption{Execution time (mins) against workload (GB) for parameter set 1.}
\label{singleNodeOnlyScala_Set1}
\end{figure*}

\begin{figure*}[t!]
$\begin{array}{rl}
    \includegraphics[width=0.48\textwidth]{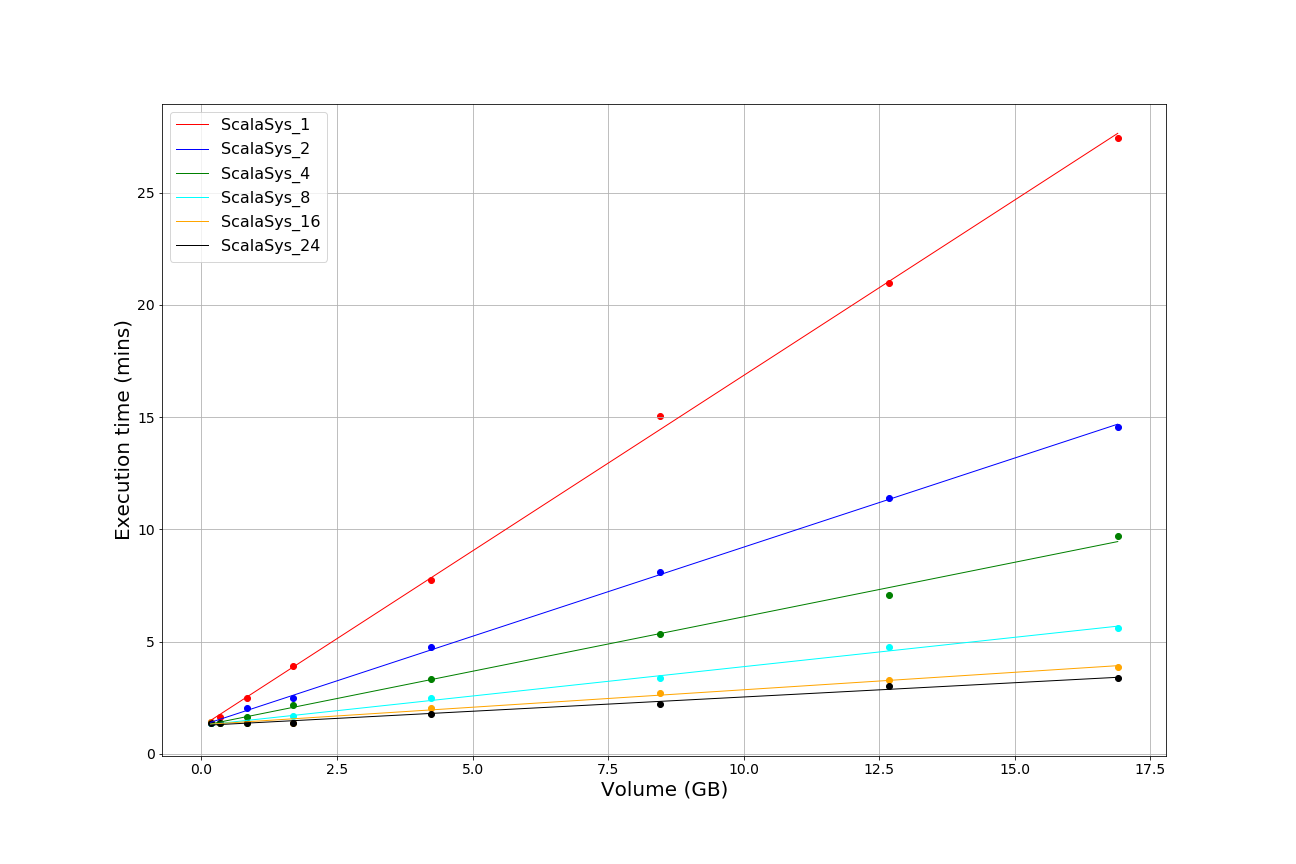} &
    \includegraphics[width=0.48\textwidth]{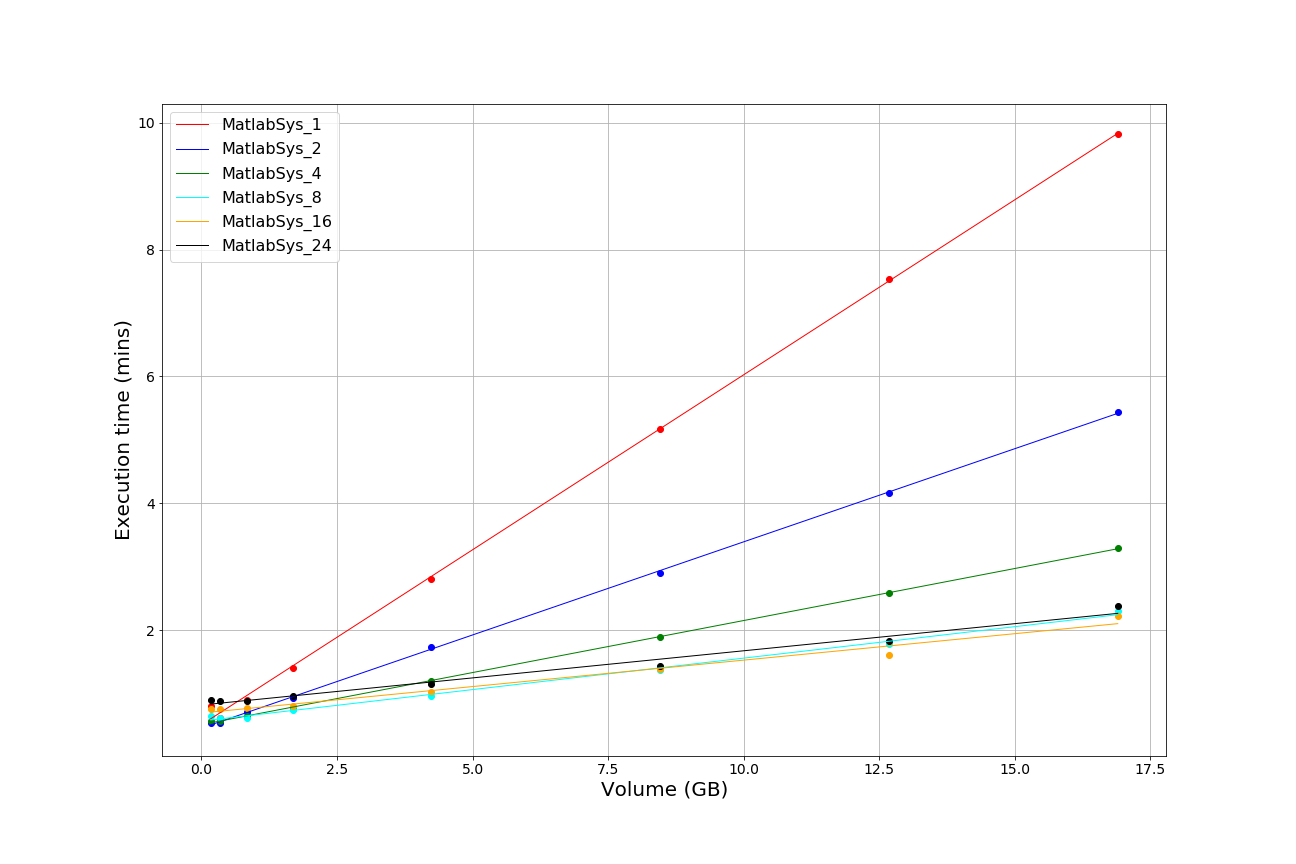}\\
    \multicolumn{2}{c}{\includegraphics[width=0.48\textwidth]{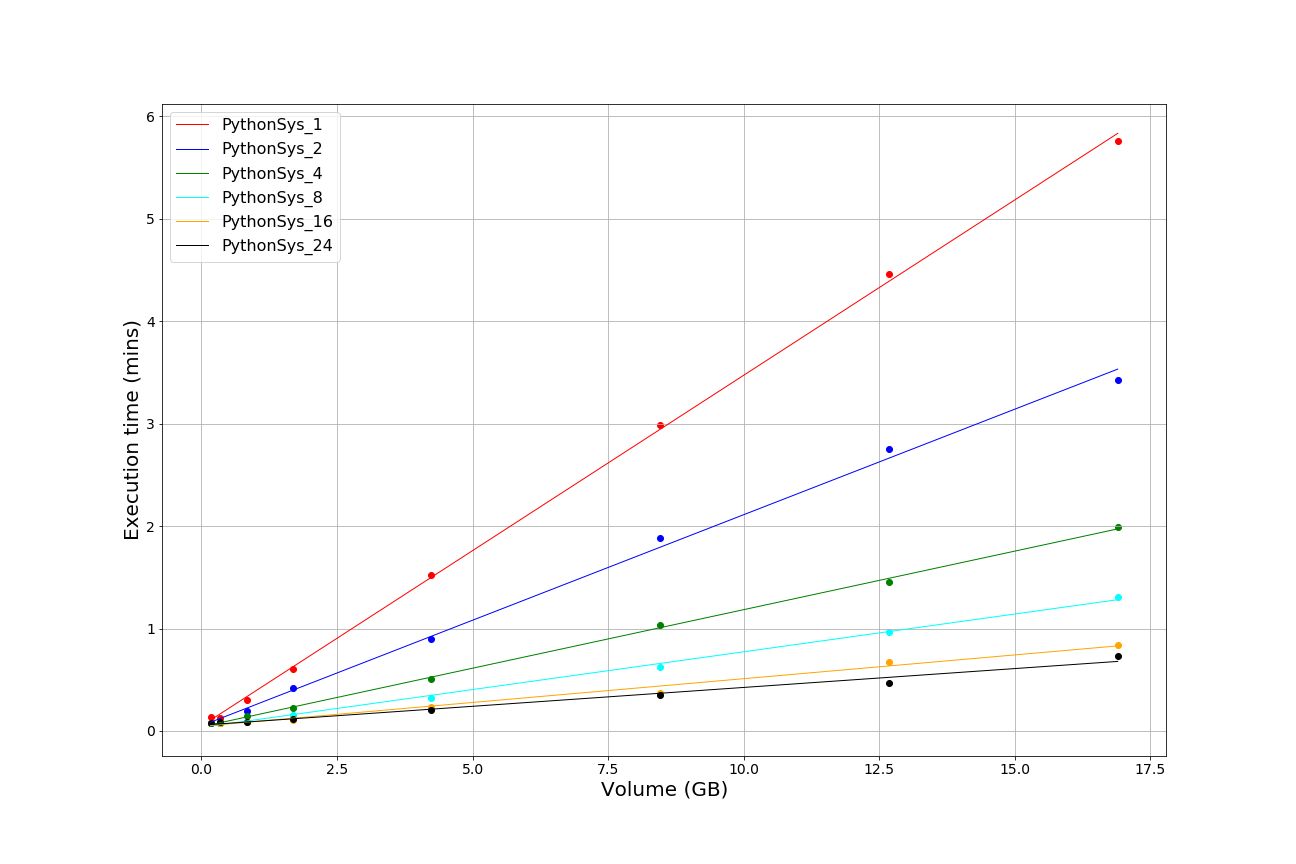}}
\end{array}$
\caption{Execution time (mins) against workload (GB) for parameter set 2.}
\label{singleNodeOnlyScala_Set2}
\end{figure*}

Figure \ref{singleNodeOnlyScala_Set1} and \ref{singleNodeOnlyScala_Set2} compare computational performance of each system using different number of threads, written after the underscore in the system names (e.g. ``MatlabSys\_2" uses two threads).

\section{Multiple node experimentation}

\begin{figure}[htbp]
\centerline{\includegraphics[width=0.8\columnwidth]{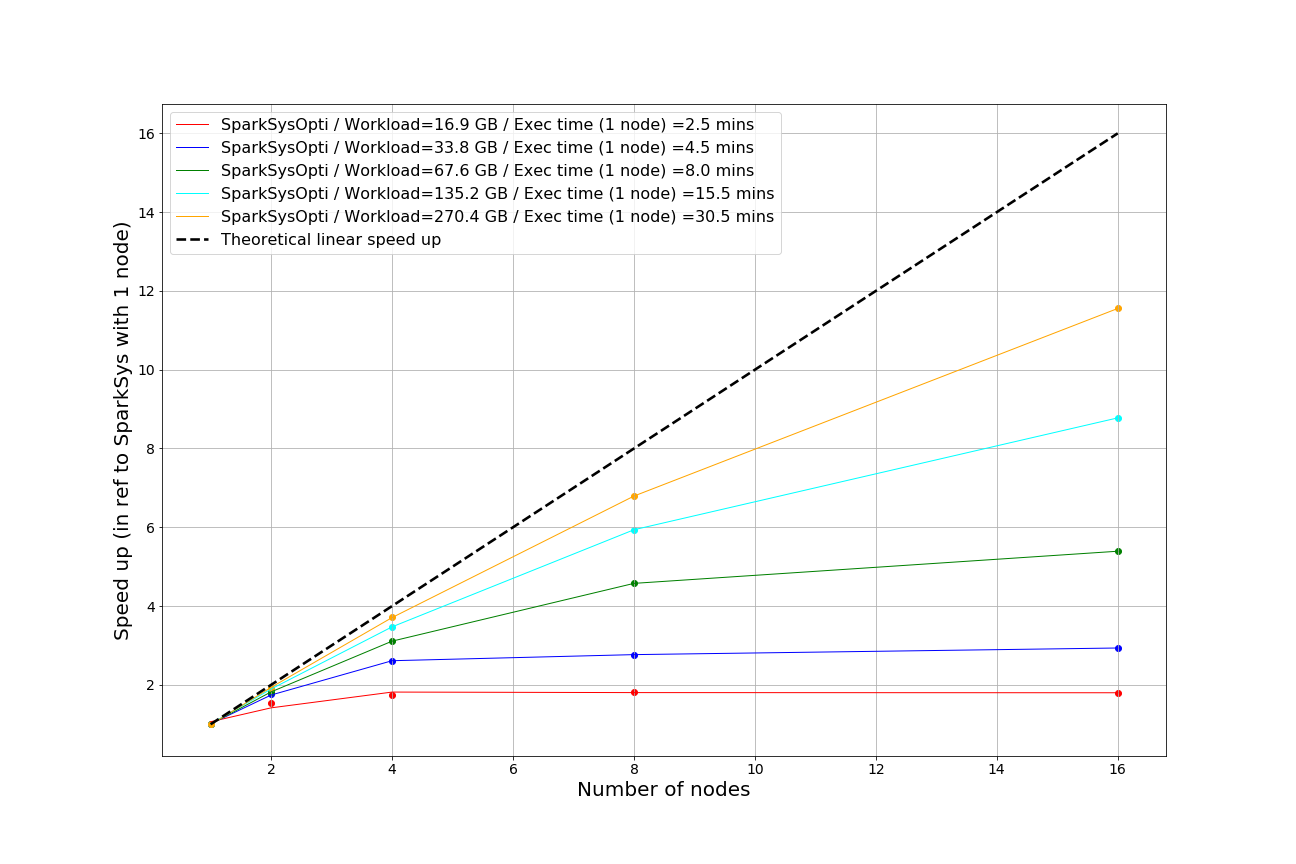}
}
\caption{Speed up metric of SparkSysOpti relatively to MatlabSys against number of cluster nodes for parameter set 1 and for different workloads (GB).}
\label{multiNode_feature_engine_benchmark_febwstw_Set1}
\end{figure}

\begin{figure}[htbp]
\centerline{\includegraphics[width=0.8\columnwidth]{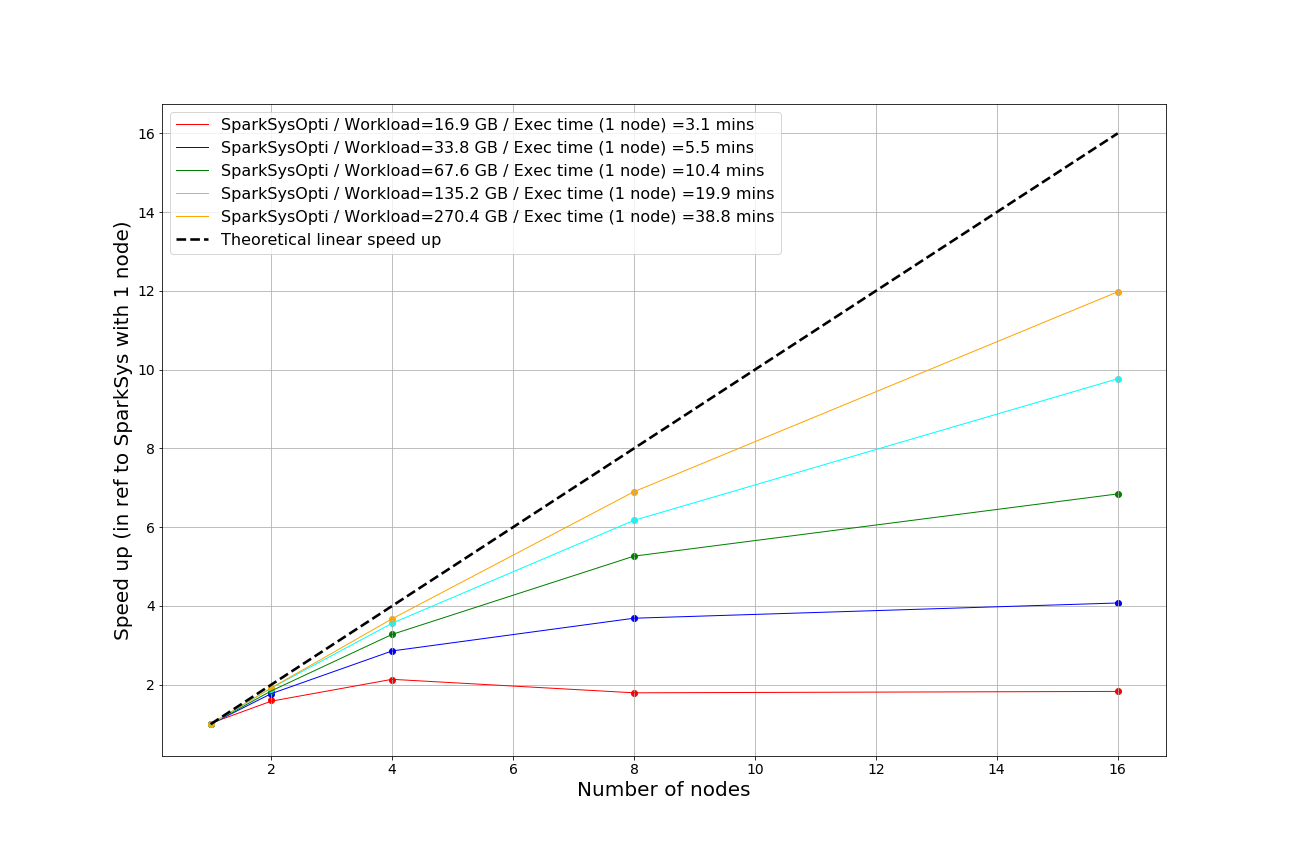}
}
\caption{Speed up metric of SparkSysOpti relatively to MatlabSys against number of cluster nodes for parameter set 2 and for different workloads (GB).}
\label{multiNode_feature_engine_benchmark_febwstw_Set2}
\end{figure}

Figures \ref{multiNode_feature_engine_benchmark_febwstw_Set1} and \ref{multiNode_feature_engine_benchmark_febwstw_Set2} represent the speed up metric of SparkSysOpti relatively to MatlabSys, as a function of the number of nodes for different workloads (in GB) for parameter sets 1 and 2, respectively. These results show that above 200 GB, an almost linear increase in speed is achieved. Indeed, as the workload increases, speed up linearizes towards the ideal case of scalability represented by the dashed black curve. For example, for parameter set 1 with a 33 GB workload, execution time only decreases by 3 when going from 1 to 4 cluster nodes, and further increasing the number of nodes up to 16 nodes does not decrease this time correspondingly. On the contrary, with a 300 GB workload, a decrease of execution time by almost 12 is observed over the increase of cluster nodes from 1 to 16. As this result has been obtained without specific optimization process, e.g. adapting the number of executors to the split length of audio file, it is very promising for further development towards a more general-purpose cloud-based analytics engine. It is noteworthy here that high-level frameworks like Hadoop / Spark highly facilitate access and democratize the use of distributed computing. On the contrary, frameworks like MPI (Message Passing Interface) would likely need more complex hand-code and fine tuning (e.g. manually setting chunking size, worker task and their synchronization), and require programming skills that are often well beyond competences of most computational scientists and researchers \cite{Dunner2017}.

As expected, our proposed system SparkSysOpti does not perform well for small-volume datasets (approximately below 250 GB in our case), as a lot of executors are made available for a small number of tasks, resulting in a lot of unused resources. One way to boost scalability here would be simply to reduce the granularity of computations, i.e. reduce the Hadoop block size so that more tasks are created and more executors work simultaneously. Similarly, running executors with too much memory often results in excessive garbage collection delays, while running tiny executors (with a single core and just enough memory needed to run a single task, for example) throws away the benefits that come from running multiple in a single JVM.

Also, the different scalability behaviours observed across the different parameter sets can easily be explained. In set 1, the system scalability is stronger than set 2 as the number of operations to be performed (i.e. FFT computations and integration over segmentation windows) per executor for each wav is more important, as we can see at the last line of table \ref{ParameterSets}, resulting in a smaller IO response time for the system (i.e. smaller waiting times for the workers).

\chapter{General discussion}

Overall, in addition to this capacity of leveraging complex analytics, we believe that Hadoop and Spark should help to reshape the big data landscape in the field of PAM research for at least three other reasons. First, Spark is able to capture fairly general computations and facilitates the implementation of iterative algorithms, e.g. used for the training algorithms of machine learning systems\footnote{Spark's machine learning library MLlib, made interoperable with NumPy}, which now play an important role in most PAM applications (e.g. for whale detection and classification, see the DCLDE workshops). It also facilitates the implementation of interactive/exploratory data analysis (i.e., the repeated database-style querying of data), especially through its SQL-compliant query capability allowing user-defined functions that leverage any general-purpose function to apply to the data columns (e.g. to rank or aggregate rows of data over a sliding window). Such computational functionalities, made here at scale with speed, are now crucial in the context of big ocean data where PAM metrics are processed conjointly with multiple heterogeneous time series from other sensors. As a result, although we focus in this work on simple FFT-based descriptor computations, we envision our Apache Hadoop/Spark big data ecosystem growing as a general-purpose analysis system useful for many different types of PAM analysis. Third, numerous efforts have been made so far to outline some best practices for PAM processing \citep{Robinson2014,Merchant2015}, in the hope of boosting standardization and interoperability. On the contrary to expensive proprietary softwares, we believe that open source software like Apache Spark will strongly contribute to this dynamic, and we would encourage computational scientists and researchers to leave behind them ``academic" codes that are too often made unreproducible, unbuildable, undocumented, unmaintained and backward incompatible.

\end{document}